\documentclass{article}
\usepackage{tikz}
\usepackage{subfigure}
\usetikzlibrary{calc}

\usepackage{amsmath}
\usepackage{amssymb}
\usepackage{amsfonts}
\usepackage{esint}
\usepackage{mathrsfs}
\usepackage{amsthm}
\usepackage{amscd}
\usepackage{longtable}
\usepackage{listings}
\usepackage{booktabs,longtable}
\usepackage{graphicx}

\usepackage{multirow}

\newcommand{\ppulp}[1]{{\mathscr #1}}

\newcommand{\ppulpm}[1]{\pmb{\ppulp{#1}}}

\newcommand{\transp}[1]{#1^{\rm T}}

\newcommand{\vect}[1]{\mathbf{#1}}

\newcommand{\matr}[1]{\mathbf{#1}}
\newcommand{\abs}[1]{\left \lvert #1 \right\rvert }

\renewcommand{\Re}[1]{\mathbb{R}\mathrm{e}\left \{#1\right\} }
\renewcommand{\Im}[1]{\mathbb{I}\mathrm{m}\left \{#1\right\} }

\newcommand{\pref}[1]{(\ref{#1})}
\newcommand{\junk}[1] {}

\def\XXint#1#2#3{{\setbox0=\hbox{$#1{#2#3}{\int}$}
\vcenter{\hbox{$#2#3$}}\kern-.5\wd0}}

\newcommand*\widebar[1]{%
  \hbox{%
    \vbox{%
      \hrule height 0.5pt 
      \kern0.3ex
      \hbox{%
        \kern-0.05em
        \ensuremath{#1}%
        \kern-0.05em
      }%
    }%
  }%
}

\begin{document}
\title{Proximity-Aware Calculation of Cable Series Impedance for Systems of Solid and Hollow Conductors}
\author{Utkarsh~R.~Patel,
        Bj\o rn~Gustavsen,
        and~Piero~Triverio
\thanks{This work was supported in part by the KPN project "Electromagnetic transients in future power systems" (ref. 207160/E20) financed by the Norwegian Research Council (RENERGI programme) and by a consortium of industry partners led by SINTEF Energy Research: DONG Energy, EdF, EirGrid, Hafslund Nett, National Grid, Nexans Norway, RTE, Siemens Wind Power, Statnett, Statkraft, and Vestas Wind Systems.}\thanks{U.~R.~Patel and P.~Triverio are with the Edward S. Rogers Sr. Department of Electrical and Computer Engineering, University of Toronto, Toronto, M5S 3G4 Canada (email: utkarsh.patel@mail.utoronto.ca, piero.triverio@utoronto.ca).}
\thanks{B.~Gustavsen is with SINTEF Energy Research, Trondheim N-7465, Norway (e-mail: bjorn.gustavsen@sintef.no).}
\\ Accepted for publication in IEEE Transactions on Power Delivery. \\ 
DOI: 10.1109/TPWRD.2014.2330994
}

\maketitle

\begin{abstract}

Wide-band cable models for the prediction of electromagnetic transients in power systems require the accurate calculation of the cable series impedance as function of frequency. A surface current approach was recently proposed for systems of round solid conductors, with inclusion of skin and proximity effects. In this paper we extend the approach to include tubular conductors, allowing to model realistic cables with tubular sheaths, armors and pipes. We also include the effect of a lossy ground. A noteworthy feature of the proposed technique is the accurate prediction of proximity effects, which can be of major importance in three-phase, pipe type, and closely-packed single-core cables. The new approach is highly efficient compared to finite elements. In the case of a cross-bonded cable system featuring three phase conductors and three screens, the proposed technique computes the required 120 frequency samples in only six seconds of CPU time.

\end{abstract}

\section{Introduction}

Insulated cables are increasingly being used in all areas of modern high-voltage power systems. As the presence of cables has a strong impact on the transient behavior of a given power system, accurate cable models should be used when analysing the system performance during transient events following circuit breaker operations, fault situations and lightning discharges. Such analyses are typically performed using suitable Electro-Magnetic Transient Programs (EMTP) \cite{EMTP, PSCAD}. As the transients may span a very broad frequency range, from a few Hz up to the MHz range, wide-band models should be used in the modeling of all relevant  system components, including cables.

The input parameters for all broadband cable models \cite{Mor99,Mar82,Nod96} are the per-unit-length matrices of series impedance and shunt admittance \cite{Pau07}. The calculation  of the series impedance is difficult, due to frequency-dependent phenomena in conductors and earth such as skin and proximity effects. In existing EMTP tools, the series impedance is obtained with analytic formulas \cite{Ame80, Wed73}, which however assume a circularly-symmetric current distribution on the conductors. This assumption becomes inaccurate in configurations combining non-coaxial arrangements and small lateral distances, like three-phase cables, pipe-type cables, and closely packed single-core cables. Here, proximity effects leads to a non-circular current distribution on conductors which is not accounted for. It has been demonstrated in numerous works  \cite{Unnur2011, Gus95} that proximity effect can significantly affect transient voltages and should therefore be taken into account. 

In a recent work~\cite{tpwrd}, the authors introduced a new method for calculating the series impedance of systems of round solid conductors which takes into account both  skin  and proximity effects. The new method, called MoM-SO, relies upon a Surface Operator (SO)~\cite{DeZ05} and the Method of Moments (MoM)~\cite{Wal08}. This surface-based approach requires only the discretization of the surface of the conductors, in contrast with volume-based approaches that mesh the entire cable cross-section, such as finite elements~\cite{Wei82,B09,Cri89} and conductor partitioning~\cite{Ame92,Com73,Dea87,Pag12,Riv02}.

The surface-based approach has been proposed in the literature for cables with conductors of rectangular~\cite{DeZ05}, triangular~\cite{Dem10}, and solid round shape~\cite{tpwrd}. Recently, an extension of the approach to hollow conductors was presented without proofs in~\cite{ipst2013}. In this paper, we present the complete derivation of the surface admittance operator for hollow conductors, and we also include the effect of ground by an approximate formulation~\cite{Gus95} where the proximity correction is added to a conventional solution which considers only skin effect. Hollow round conductors are useful in modeling  realistic cable systems with tubular sheaths, armors,  and pipes. The extended MoM-SO method is validated against a finite element (FEM) computation.  Finally, we demonstrate the complete procedure with the modeling and simulation of  a cross-bonded cable system involving three closely-packed single-core cables.

\section{Problem Statement}

We consider a cable made by $P$ round conductors oriented along the $z$-axis and surrounded by a lossless medium of permittivity $\varepsilon_{o}$ and permeability $\mu_o$. The cross section of each conductor can be either solid, as in the left panel of Fig.~\ref{fig:solid}, or hollow, as in the left panel of Fig.~\ref{fig:hollow}. We denote the outer radius of the $p$-th conductor with $a_p$. If the conductor is hollow, we denote its inner radius with $\tilde{a}_p$. The conductors have conductivity $\sigma$, permittivity $\varepsilon$, and permeability $\mu$.

From the geometry of the cable, we aim to compute the p.u.l.\footnote{per unit length.} resistance  $\ppulpm{R}(\omega)$  and  inductance  $\ppulpm{L}(\omega)$ matrices which appear in the Telegrapher's equation~\cite{Pau07}
\begin{equation}
\frac{\partial \vect{V}}{\partial z} = - \left [\ppulpm{R}(\omega) + j\omega\ppulpm{L}(\omega)\right ] \vect{I}\,,
\label{eq:pteleg}
\end{equation}
where we collect the potential $V_p$ and the current $I_p$ of each conductor in the column vectors $\vect{V} = \begin{bmatrix} V_1 & \hdots & V_P \end{bmatrix}^T$,  and $\vect{I} = \begin{bmatrix} I_1 & \hdots & I_{P} \end{bmatrix}^T$, respectively.

\section{Surface Admittance Formulation}
\label{sec:SurfaceAdmittance}
We calculate the p.u.l. resistance and inductance of the cable with the surface method introduced in~\cite{DeZ05}, where a surface admittance operator is used to replace all conductors with an equivalent current on their boundary. We first present the surface admittance representation focusing on a single solid or hollow conductor. Then, in Sec.~\ref{sec:samultiple}, the representation is extended to all conductors in the cable, and used in Sec.~\ref{sec:impedancecomputation} to compute the cable parameters.

\subsection{Surface Admittance Operator for a Solid Conductor}
\label{sec:saosolid}

\begin{figure}
\centering    
\subfigure{
\begin{tikzpicture}
\draw [lightgray,fill=lightgray] (0,0) rectangle (3.75,3);
\node [above] at (3,0) {$\varepsilon_{o}, \mu_o$};

\draw [black,fill=white] (1.7,1.5) circle [radius=1.2] node [below] {$\varepsilon, \mu, \sigma$};
\draw [dotted] (1.7,1.5) -- (3.2,1.5);
\draw [dotted] (1.7,1.5) --+ (45:1.5);
\draw [->] (3.2,1.5) arc (0:45:1.5);
\node [right] at (3.1,2.05) {$\theta$};
\draw [thick, ->] (1.7,1.5) -- +(45:1.2);
\node [above] at (1.9,1.7) {$a_p$};
\end{tikzpicture}
}
\subfigure{
\begin{tikzpicture}
\draw [lightgray,fill=lightgray] (0,0) rectangle (3.75,3);
\node [above] at (3,0) {$\varepsilon_{o}, \mu_o$};
\draw [black, dashed] (1.7,1.5) circle [radius=1.2];

\node [right] at (.7,.3) {$c_p$};

\foreach \a in {20, 80,...,320} {
      \draw ({1.7+1.2*cos(\a)},{1.5+1.2*sin(\a)}) circle [radius=.1];
      \draw [fill=black] ({1.7+1.2*cos(\a)},{1.5+1.2*sin(\a)}) circle [radius=.04];
}

\node [above left] at (1.20,2.35) {$J_s^{(p)}(\theta)$};
\end{tikzpicture}
}
\caption{Application of the equivalence theorem to a solid round conductor. The conductor (left panel) is replaced by the surrounding medium and an equivalent current $J_s^{(p)}(\theta)$ on its surface (right panel). The conductor radius is denoted with $a_p$.}
\label{fig:solid}
\end{figure}
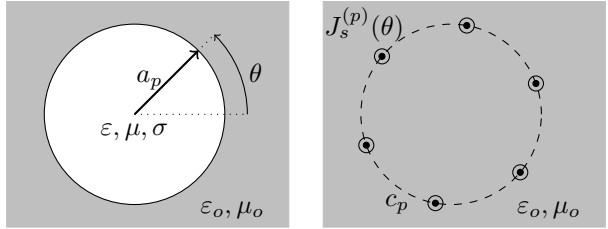

We let conductor $p$ be solid with the cross section shown in the left panel of Fig.~\ref{fig:solid}. Using cylindrical coordinates, we trace the boundary $c_p$ of the conductor with the position vector 
\begin{equation}
\vect{r}_p(\theta) = \left (x_p + a_p \cos \theta  \right ) { \hat{\vect{x}}} +
			\left (y_p + a_p \sin \theta  \right ) { \hat{\vect{y}}}\,,
\label{eq:rtheta}
\end{equation}
where $(x_p, y_p)$ is the position of the conductor's center and $\hat{\vect{x}}, \hat{\vect{y}}$ denote the unit vectors along the $x$- and $y$-axis, respectively. The longitudinal component of the electric field on the conductor boundary is denoted with $E_z^{(p)}(\theta)$, and is expanded in truncated Fourier series
\begin{equation}
E_z^{(p)}(\theta) = \sum_{n=-N_p}^{N_p} E_{n}^{(p)} e^{j n \theta}\,,
\label{eq:truncatedbound1}
\end{equation}
where $N_p$ controls the number of harmonics taken into account. The case $N_p =0$ corresponds to assuming a circularly-symmetric current distribution in the conductor. This assumption, made by analytic formulas used in existing EMTP tools, is accurate only for well-separated conductors. When spacing is comparable to the conductors size, proximity effects become significant and lead to a non-uniform field distribution that calls for $N_p > 0$. Numerical tests~\cite{tpwrd} show that a $N_p$ of 3 or 4 is sufficient in most cases. The choice of the parameter $N_p$ can be performed automatically~\cite{Spi2013}, and will not be discussed here.

Given the field~\pref{eq:truncatedbound1} on the boundary, the electric field inside the conductor ${\ppulp E}_z^{(p)}(\rho, \theta)$ can be found by solving the Helmholtz equation~\cite{Poz04} and reads
\begin{equation}
{\ppulp E}_z^{(p)}(\rho, \theta) = \sum_{n=-N_p}^{N_p} \frac{E_n^{(p)}}{{\cal J}_{|n|}(ka_p)} {\cal J}_{|n|}(k\rho)  e^{j n \theta}\,, 
\label{eq:Efield1}
\end{equation}
where 
\begin{align}
k & = \sqrt{ \omega \mu (\omega \varepsilon -j \sigma)}\,
\label{eq:k}
\end{align}
is the wavenumber inside the conductor, $\rho \in [0,a_p]$ is the radial coordinate, and ${\cal J}_{|n|}(.)$ is the Bessel function of the first kind of order $|n|$~\cite{Abr64}.

We next use the equivalence theorem~\cite{Har61,DeZ05} to replace the conductor with the surrounding medium. On its boundary $c_p$, we introduce an equivalent surface current density $J_s^{(p)}(\theta)$ in order to keep the electromagnetic field \emph{outside} the conductor unchanged. This transformation is illustrated in  Fig.~\ref{fig:solid}. The value of the equivalent current $J_s^{(p)}(\theta)$ is given by the equivalence theorem~\cite{Har61,DeZ05} as
\begin{equation}
J_s^{(p)}(\theta) = H_t^{(p)}(a_p^{-}, \theta) - \widebar{H}^{(p)}_t(a_p^{-}, \theta)  \,,
\label{eq:Jdef}
\end{equation}
where $H_t^{(p)}(a_p^{-}, \theta)$ and $\widebar{H}_t^{(p)}(a_p^{-}, \theta)$ denote the magnetic field tangential to the conductor boundary respectively \emph{before} and \emph{after} the application of the equivalence theorem. Both fields are evaluated just inside the conductor boundary (i.e. for $\rho = a_p^{-}$).

The magnetic fields in~\pref{eq:Jdef} can be related to the longitudinal electric field as~\cite{DeZ05}
\begin{align}
H_{t}^{(p)} (a^-_p, \theta) &= \frac{1}{j\omega \mu} \frac{\partial {\ppulp E}_z^{(p)}}{\partial \rho} \bigg |_{\rho = a^-_p} \label{eq:HErel1} \,, \\
\widebar{H}_{t}^{(p)} (a^-_p, \theta) &= \frac{1}{j\omega \mu_o} \frac{\partial \widebar{{\ppulp E}}_z^{(p)}}{\partial \rho} \bigg |_{\rho = a^-_p}\, \,,
\label{eq:HErel2}
\end{align}
where $\widebar{{\ppulp E}}_z^{(p)}(\rho, \theta)$ is the electric field inside $c_p$ after application of the equivalence theorem. This field can be written as
\begin{equation}
\widebar{{\ppulp E}}_z^{(p)}(\rho, \theta) = \sum_{n=-N_p}^{N_p} \frac{E_n^{(p)}}{{\cal J}_{|n|}(k_o a_p) } {\cal J}_{|n|}(k_o\rho)  e^{j n \theta}\,,
\label{eq:Efieldbar1}
\end{equation}
which is~\pref{eq:Efield1} where $k$ has been replaced with the wavenumber of the surrounding medium
\begin{align}
k_{o} & = \omega \sqrt{\mu_o \varepsilon_{o}}\,.
\label{eq:kout}
\end{align}
By substituting~\pref{eq:HErel1} and~\pref{eq:HErel2} into~\pref{eq:Jdef}, we obtain
\begin{equation}
J_s^{(p)}(\theta) = \frac{1}{j\omega} \left [ \frac{1}{\mu } \frac{\partial {\ppulp E}_z^{(p)}}{\partial \rho}  - \frac{1}{\mu_o} \frac{\partial \widebar{{\ppulp E}}_z^{(p)}}{\partial \rho} \right ]_{\rho = a_p^-}\,. \label{eq:saosolids}
\end{equation}
Equation~\pref{eq:saosolids} defines a surface admittance operator that relates the equivalent current density to the electric field on the conductor boundary. If we adopt for the equivalent current density ${J}_s^{(p)}(\theta)$ a truncated Fourier expansion analogous to~\pref{eq:truncatedbound1}
\begin{align}
{J}_s^{(p)}(\theta) &= \frac{1}{2\pi a_p}\sum_{n=-N_p}^{N_p} J^{(p)}_{n} e^{j n \theta}\,,
\end{align}
we can conveniently express the surface admittance operator~\pref{eq:saosolids} in terms of the Fourier coefficients $J_{n}^{(p)}$ and $E_{n}^{(p)}$ as~\cite{DeZ05}
\begin{equation}
J_{n}^{(p)} =  Y_n^{(p)} E_{n}^{(p)}\,,
\label{eq:JCoeff}
\end{equation}
where
\begin{equation}
Y_n^{(p)} = \frac{2\pi}{j\omega } \biggl [ \frac{k a_p {\cal J}_{|n|}'(ka_p)}{\mu {\cal J}_{|n|}(ka_p)} - \frac{k_{o} a_p {\cal J}_{|n|}'(k_{o}a_p)}{\mu_o {\cal J}_{|n|}(k_{o}a_p)}  \biggl ]\,,
\label{eq:Yn}
\end{equation}
and where ${\cal J}_{|n|}'(.)$ is the derivative of ${\cal J}_{|n|}(.)$.
Before exploiting~\pref{eq:JCoeff} for the computation of the series impedance, we extend the surface operator to hollow conductors.

\subsection{Surface Admittance Operator for a Hollow Conductor}

\begin{figure}
\centering   
\subfigure{
\begin{tikzpicture}
\draw [lightgray,fill=lightgray] (0,0) rectangle (3.75,3);
\node [above] at (3,0) {$\varepsilon_{o}, \mu_o$};
\draw [black,fill=white] (1.7,1.5) circle [radius=1.2] node [below] {};
\node [above] at (1.7,0.3) {$\varepsilon, \mu, \sigma$};
\draw [black,fill=lightgray] (1.7,1.5) circle [radius=0.7] node [below] {};

\draw [thick, ->] (1.7,1.5) -- (0.851,2.349);
\node [left] at (1.5755,1.6245) {$a_p$};
\draw [thick, ->] (1.7,1.5) -- (2.195,2.0);
\node [below] at (2.095,1.85) {$\tilde{a}_p$};
\end{tikzpicture}
}
\subfigure{
\begin{tikzpicture}
\draw [lightgray,fill=lightgray] (0,0) rectangle (3.75,3);
\node [above] at (3,0) {$\varepsilon_{o}, \mu_o$};
\draw [black, dashed] (1.7,1.5) circle [radius=1.2];
\draw [black, dashed] (1.7,1.5) circle [radius=0.7];
\foreach \a in {20, 80,...,320} {
      \draw ({1.7+1.2*cos(\a)},{1.5+1.2*sin(\a)}) circle [radius=.1];
      \draw [fill=black] ({1.7+1.2*cos(\a)},{1.5+1.2*sin(\a)}) circle [radius=.04];
}
\foreach \a in {20, 80,...,320} {
      \draw ({1.7+0.7*cos(\a)},{1.5+0.7*sin(\a)}) circle [radius=.1];
      \draw [fill=black] ({1.7+0.7*cos(\a)},{1.5+0.7*sin(\a)}) circle [radius=.04];
}
\node [above] at (0.6,2.35) {$J_s^{(p)}(\theta)$};
\node [right] at (2.45,2.45) {$c_p$};
\node [right] at (2.05,2.15) {$\tilde{c}_p$};
\node [above right] at (1.0,1.2) {$\widetilde{J}_s^{(p)}(\theta)$};
\end{tikzpicture}

}
\caption{Application of the equivalence theorem to a hollow conductor. 
The actual conductor, shown in the left panel, is replaced by the surrounding medium and equivalent currents $\widetilde{J}_s^{(p)}(\theta)$ and $J_s^{(p)}(\theta)$ are introduced on the inner and outer surface of the conductor (right panel). The inner and outer radius are denoted with $\tilde{a}_p$ and $a_p$, respectively.}
\label{fig:hollow}
\end{figure}
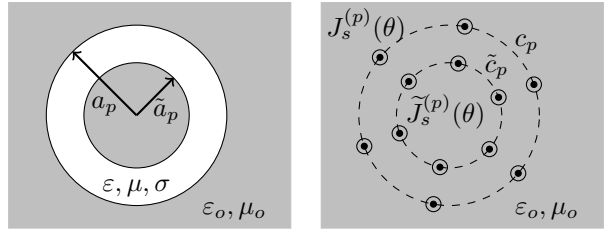

We now consider a hollow conductor with the cross section depicted in the left panel of Fig.~\ref{fig:hollow}. In addition to the outer boundary $c_p$, we now have an inner boundary, denoted with $\tilde{c}_p$. The outer boundary is traced by the position vector~\pref{eq:rtheta}, while the inner boundary is traced by
\begin{equation}
\tilde{\vect{r}}_p(\theta) = \left (x_p + \tilde{a}_p \cos \theta  \right ) { \hat{\vect{x}}} +
			\left (y_p + \tilde{a}_p \sin \theta  \right ) { \hat{\vect{y}}}\,.
\label{eq:ritheta}
\end{equation}
The electric field on the inner and outer boundaries are denoted with $\widetilde{E}_z^{(p)}(\theta)$ and ${E}_z^{(p)}(\theta)$, respectively, and are approximated by Fourier series as
\begin{align}
{E}_z^{(p)}(\theta) &= \sum_{n=-N_p}^{N_p} E_{n}^{(p)} e^{j n \theta}\,  \label{eq:truncatedhbound1} \,,\\
\widetilde{E}_z^{(p)}(\theta) &= \sum_{n=-N_p}^{N_p} \widetilde{E}_{n}^{(p)} e^{j n \theta}\, \,.
\label{eq:truncatedhbound2}
\end{align}
Given the boundary conditions~\pref{eq:truncatedhbound1} and~\pref{eq:truncatedhbound2}, the electric field $\ppulp{E}^{(p)}_z(\rho, \theta)$  inside the conductor can be found by solving the Helmholtz equation in a hollow region~\cite{Che89}, and reads
\begin{align}
\ppulp{E}^{(p)}_z(\rho,\theta) = \sum_{n=-N_p}^{N_p} \bigg ( &C_{n}(k) {\cal H}_{|n|}(k \rho) + \nonumber \\
& D_{n}(k) {\cal K}_{|n|}(k \rho) \bigg)  e^{jn\theta} \,,
\label{eq:Ehollow1}
\end{align}
where the constants $C_{n}(k)$ and $D_{n}(k)$ are found from the boundary conditions. Imposing~\pref{eq:truncatedhbound1} and~\pref{eq:truncatedhbound2} on the two boundaries, we obtain
\begin{align}
	C_{n}(k) &= \frac{E_n^{(p)}{\cal K}_{|n|}(k \tilde{a}_p) - \widetilde{E}_n^{(p)}{\cal K}_{|n|}(k a_p)}{m_n(ka_p,k\tilde{a}_p)} \,, \label{eq:cn}\\
	D_{n}(k) &= \frac{\widetilde{E}_n^{(p)}{\cal H}_{|n|}(k a_p) - E_n^{(p)}{\cal H}_{|n|}(k \tilde{a}_p)}{m_n(ka_p,k \tilde{a}_p)} \,,
\label{eq:dn}
\end{align}
where
\begin{align}
	m_n(\alpha,\beta) & = {\cal H}_{|n|}(\alpha) {\cal K}_{|n|}(\beta) - {\cal H}_{|n|}(\beta) {\cal K}_{|n|}(\alpha)\,.
\end{align}
In these formulas, ${\cal H}_{|n|} \left( .\right)$ and ${\cal K}_{|n|} \left( .\right)$ denote the Hankel functions of order $|n|$ of, respectively, the first and second kind~\cite{Abr64}. 

In analogy with what done for a solid conductor, we replace the conductor with the surrounding medium as shown in the right panel of Fig.~\ref{fig:hollow}. Now, two equivalent current densities must be introduced, one on the outer boundary denoted with ${J}_s^{(p)}(\theta)$, and one on the inner boundary denoted with $\widetilde{J}_s^{(p)}(\theta)$. Both current densities are approximated by truncated Fourier series
\begin{align}
	{{J}}_s^{(p)}(\theta)  = \frac{1}{2\pi {a}_p}\sum_{n=-N_p}^{N_p} J_{n}^{(p)} e^{j n \theta}\,, \label{eq:Jout} \\
	\widetilde{{J}}_s^{(p)}(\theta)  = \frac{1}{2\pi \tilde{a}_p}\sum_{n=-N_p}^{N_p} \widetilde{J}_{n}^{(p)} e^{j n \theta}\,. \label{eq:Jin}
\end{align} 
Using the equivalence theorem~\cite{Har61}, we size the equivalent currents in order to preserve the original electric field both inside the cavity ($\rho < \tilde{a}_p$) and beyond the outer boundary ($\rho > a_p$). With a derivation analogous to the one presented in Sec.~\ref{sec:saosolid} for a solid conductor, we obtain the following expression for the equivalent currents
\begin{equation}
J_s^{(p)}(\theta) = \frac{1}{j\omega} \left [ \frac{1}{\mu } \frac{\partial {\ppulp E}_z^{(p)}}{\partial \rho}  - \frac{1}{\mu_o} \frac{\partial \widebar{{\ppulp E}}_z^{(p)}}{\partial \rho} \right ]_{\rho = a_p^-}\,, \label{eq:saohollow1}
\end{equation}
and
\begin{equation}
\widetilde{J}_s^{(p)}(\theta) = \frac{1}{j\omega} \left [ \frac{1}{\mu_o} \frac{\partial \widebar{{\ppulp E}}_z^{(p)}}{\partial \rho} - \frac{1}{\mu } \frac{\partial {\ppulp E}_z^{(p)}}{\partial \rho} \right ]_{\rho = \tilde{a}_p^+}\,, \label{eq:saohollow2}
\end{equation}

where $\widebar{\ppulp{E}}^{(p)}_z(\rho, \theta)$ is the electric field inside the conductor \emph{after} the application of the equivalence theorem. Its value is given by~\pref{eq:Ehollow1} with $k$ replaced by the wavenumber of the surrounding medium~\pref{eq:kout}.
Equations~\pref{eq:saohollow1} and~\pref{eq:saohollow2} define the surface admittance operator for a hollow conductor. This result is a generalization of the surface admittance operator given in~\cite{DeZ05} for solid conductors.

We now rewrite the operator~\pref{eq:saohollow1}-\pref{eq:saohollow2} in terms of the Fourier coefficients $J_n^{(p)}$, $\widetilde{J}_n^{(p)}$, $E_n^{(p)}$ and $\widetilde{E}_n^{(p)}$. Using~\pref{eq:Ehollow1}, we can write the derivatives in~\pref{eq:saohollow1} and~\pref{eq:saohollow2} as
\begin{align}
& \frac{\partial {\ppulp E}^{(p)}_z}{\partial \rho} = \!\!\! \sum_{n=-N_p}^{N_p} \!\!\! \left [ C_{n}(k) {\cal H'}_{|n|}(k \rho) + D_{n}(k) {\cal K'}_{|n|}(k \rho) \right ]  k e^{jn\theta}  \label{eq:dE2} \\
& \frac{\partial \widebar{{\ppulp E}}^{(p)}_z}{\partial \rho} = \!\!\! \sum_{n=-N_p}^{N_p} \!\!\! \left [ C_{n}(k_o) {\cal H'}_{|n|}(k_o \rho) + D_{n}(k_o) {\cal K'}_{|n|}(k_o \rho) \right ]  k_o e^{jn\theta}
\label{eq:dE1}
\end{align}
where ${\cal H}'_{|n|}(.)$ and ${\cal K}'_{|n|}(.)$ are the derivatives of the Hankel functions ${\cal H}_{|n|}(.)$ and ${\cal K}_{|n|}(.)$, respectively. Substituting~\pref{eq:dE2}-\pref{eq:dE1} into~\pref{eq:saohollow1}-\pref{eq:saohollow2}, we can finally write the surface admittance operator in terms of the unknown Fourier coefficients as
\begin{equation}
	\begin{bmatrix}
		\widetilde{J}^{(p)}_{n} \\
		J^{(p)}_{n}
	\end{bmatrix} =
	\matr{Y}_n^{(p)}
	\begin{bmatrix}
		\widetilde{E}^{(p)}_{n} \\
		E^{(p)}_{n}
	\end{bmatrix}\,.
	\label{eq:Yhollow}
\end{equation}
In this equation,
\begin{equation}
	\matr{Y}_n^{(p)} =
	\begin{bmatrix}
		Y^{}_{11,n} & Y^{}_{12,n} \\
		Y^{}_{21,n} & Y^{}_{22,n}
	\end{bmatrix}
	\label{eq:Yhollow2}
\end{equation}
is a $2 \times 2$ matrix which generalizes~\pref{eq:Yn} to the hollow conductor case. The matrix entries are given by
\begin{align*}
Y^{}_{n,11} & =  \frac{2\pi}{j \omega} \left [ \frac{\chi_n(k a_p, k \tilde{a}_p)}{m_n(k a_p, k \tilde{a}_p) \mu} 
	- \frac{\chi_n(k_{o} a_p, k_{o} \tilde{a}_p)}{m_n(k_{o} a_p, k_{o} \tilde{a}_p) \mu_o}  \right ]\\
Y^{}_{12,n} & = \frac{2\pi}{j \omega} \left [ \frac{\chi_n(k_{o} \tilde{a}_p, k_{o} \tilde{a}_p)}{m_n(k_{o} a_p, k_{o} \tilde{a}_p) \mu_o} 
	-  \frac{\chi_n(k \tilde{a}_p, k \tilde{a}_p)}{m_n(k a_p, k \tilde{a}_p) \mu} \right ]\\
Y^{}_{21,n} & = \frac{2\pi}{j \omega} \left [ \frac{\chi_n(k_{o} a_p, k_{o} a_p)}{m_n(k_{o} a_p, k_{o} \tilde{a}_p) \mu_o} 
	-  \frac{\chi_n(k a_p, k a_p)}{m_n(k a_p, k \tilde{a}_p) \mu} \right ]\\
Y^{}_{22,n} & =  \frac{2\pi}{j \omega} \left [ \frac{\chi_n(k \tilde{a}_p, k a_p)}{ m_n(k a_p, k \tilde{a}_p) \mu} 
	- \frac{\chi_n(k_{o} \tilde{a}_p, k_{o} a_p)}{ m_n(k_{o} a_p, k_{o} \tilde{a}_p) \mu_o}  \right ]
\end{align*}
with
\begin{equation*}
\chi_n(\alpha,\beta) = \beta \left [  {\cal H}'_{|n|}(\beta) {\cal K}_{|n|}(\alpha) - {\cal H}_{|n|}(\alpha) {\cal K}'_{|n|}(\beta)    \right ] \,.
\end{equation*}

\subsection{Surface Admittance Operator for Multiple Conductors}
\label{sec:samultiple}

We now apply the surface admittance operator to all conductors in the cable, introducing equivalent currents on their boundaries. In order to simplify the notation for upcoming formulas, we gather all Fourier coefficients related to conductor $p$ in two column vectors $\matr{E}^{(p)}$ and $\matr{J}^{(p)}$. If conductor $p$ is solid, we let
\begin{align}
\vect{E}^{(p)} &=
\begin{bmatrix}
E^{(p)}_{-N_p} & \hdots & E^{(p)}_{0} & \hdots & E^{(p)}_{N_p}
\end{bmatrix} \,, \\	
\vect{J}^{(p)} &=
\begin{bmatrix}
J^{(p)}_{-N_p} & \hdots & J^{(p)}_{0} & \hdots & J^{(p)}_{N_p}
\end{bmatrix} \,.
\end{align}
If conductor $p$ is hollow, we set
\begin{align}
\vect{E}^{(p)} &=
\begin{bmatrix}
\widetilde{E}^{(p)}_{-N_p} &  \hdots & \widetilde{E}^{(p)}_{N_p} & E^{(p)}_{-N_p} &  \hdots & E^{(p)}_{N_p}
\end{bmatrix}	\,, \\
\vect{J}^{(p)} &=
\begin{bmatrix}
\widetilde{J}^{(p)}_{-N_p} &  \hdots & \widetilde{J}^{(p)}_{N_p} & J^{(p)}_{-N_p} &  \hdots & J^{(p)}_{N_p}
\end{bmatrix}\,.
\end{align}
Furthermore, all electric field and current coefficients are collected in the global vectors of unknowns
\begin{align}
\matr{E} &= 
\begin{bmatrix}
\vect{E}^{(1)} & \vect{E}^{(2)} & \cdots & \vect{E}^{(p)} & \cdots & \vect{E}^{(P)}
\end{bmatrix}^T \,, \label{eq:Emat}\\
\matr{J} &= 
\begin{bmatrix}
\vect{J}^{(1)} & \vect{J}^{(2)} & \cdots & \vect{J}^{(p)} & \cdots & \vect{J}^{(P)}
\end{bmatrix}^T.
\label{eq:Jmat}
\end{align}
The current coefficients in~\pref{eq:Jmat} are related to the electric field coefficients~\pref{eq:Emat} by the surface admittance operators~\pref{eq:JCoeff} and~\pref{eq:Yhollow}. All these relations can be summarized in matrix form as
\begin{equation}
\matr{J} = \matr{Y}_s \matr{E}\,,
\label{eq:sao}
\end{equation}
where the block diagonal matrix $\matr{Y}_s$ can be interpreted as the surface admittance operator of the whole system of conductors.

\section{Impedance Computation}
\label{sec:impedancecomputation}
\subsection{Electric Field Integral Equation}

The surface admittance operator describes the field-current relation imposed by the conductors. The effect of the surrounding medium is instead modelled with  the electric field integral equation~\cite{Bal05,DeZ05}
\begin{equation}
\ppulp{E}_z(\vect{r}) = -j\omega A_z(\vect{r}) - \frac{\partial V}{\partial z} \,,
\label{eq:diffefie}
\end{equation}
where $V$ is the scalar potential and 
\begin{equation}
A_z(\vect{r}) = -\mu_o \int J_s(\vect{r}') G(\vect{r}, \vect{r}') d\vect{r}'
\label{eq:defA}
\end{equation}
is the $z$-component of the vector potential, which is obtained by superimposing the effect of the equivalent currents through the Green's function $G(\vect{r}, \vect{r}')$. Since, after the equivalence theorem has been applied to all conductors, the entire medium has become homogeneous, $G(\vect{r}, \vect{r '})$ is simply the Green's function of a two-dimensional infinite space~\cite{Har61}
\begin{equation}
G(\vect{r},\vect{r}') = \frac{1}{2\pi} \ln \abs{\vect{r}-\vect{r}'} \,.
\label{eq:green}
\end{equation}
We can write the vector potential $A_z(\vect{r})$ as
\begin{equation}
	A_z(\vect{r}) = \sum_{q = 1}^{P} A_q(\vect{r}) \,,
\end{equation}
where $A_q(\vect{r})$ is the contribution of the current that replaced conductor $q$. If conductor $q$ is solid we have
\begin{equation}
A_q(\vect{r}) = 
-\mu_o \int_0^{2\pi} J^{(q)}_{s}(\theta') G \left ( \vect{r},\vect{r}_q(\theta') \right ) a_q d\theta '  \,,
\end{equation}
while if conductor $q$ is hollow we have
\begin{align}
A_q(\vect{r}) =& 
-\mu_o \int_0^{2\pi} J^{(q)}_{s}(\theta') G \left ( \vect{r},\vect{r}_q(\theta') \right )  a_q d\theta '\\
&-\mu_o \int_0^{2\pi} \widetilde{J}^{(q)}_{s}(\theta') G \left ( \vect{r},\vect{\widetilde{r}}_q(\theta') \right ) \tilde{a}_q  d\theta '  \, \nonumber
\end{align}
since we have to superimpose the effect of the equivalent current on both the inner and outer contours.

When $\vect{r}$ belongs to the outer boundary of conductor $p$, we can rewrite~\pref{eq:diffefie} as
\begin{equation}
E_z^{(p)}(\theta) = -j\omega A_z(\vect{r}_p(\theta)) + \sum_{q=1}^P \left [ \ppulpm{R}_{pq}(\omega)+j\omega \ppulpm{L}_{pq}(\omega) \right ] I_q \,,
\label{eq:diffefie2}
\end{equation}
where $\ppulp{E}_z(\vect{r})$ has been replaced by its Fourier expansion~\pref{eq:truncatedbound1}, and the term $\frac{\partial V}{\partial z}$ has been written through~\pref{eq:pteleg}. The symbols $\ppulpm{R}_{pq}(\omega)$ and $\ppulpm{L}_{pq}(\omega)$ represent the $(p,q)$ entry of the matrices $\ppulpm{R}(\omega)$ and $\ppulpm{L}(\omega)$, respectively. Equation~\pref{eq:diffefie2} is written for all conductors, both solid and hollow. In addition, if $p$ is a hollow conductor, we also evaluate~\pref{eq:diffefie} on the inner boundary $\tilde{c}_p$, obtaining
\begin{equation}
\widetilde{E}_z^{(p)}(\theta) = -j\omega A_z(\tilde{\vect{r}}_p(\theta)) + \sum_{q=1}^P \left [ \ppulpm{R}_{pq}(\omega)+j\omega \ppulpm{L}_{pq}(\omega) \right ] I_q \,.
\label{eq:diffefie3}
\end{equation}

The integral equations~\pref{eq:diffefie2} and~\pref{eq:diffefie3} can be solved numerically with the method of moments~\cite{Wal08}, using the Fourier expansions~\pref{eq:truncatedhbound1}, \pref{eq:truncatedhbound2}, \pref{eq:Jout} and \pref{eq:Jin} for the unknown fields and currents. This process was presented in~\cite{tpwrd} and is here omitted due to the limited space. It finally leads to
\begin{equation}
\vect{E} = j\omega \mu_o \matr{G} \vect{J} + \matr{U} \left [\ppulpm{R}(\omega) + j \omega\ppulpm{L}(\omega)\right ] \matr{U}^T \vect{J}\,,
\label{eq:mom}
\end{equation}
which is an algebraic approximation of~\pref{eq:diffefie2} and~\pref{eq:diffefie3}.
This system of equations relates the Fourier coefficients $\vect{E}$ and $\vect{J}$ of the unknowns, and combined with~\pref{eq:sao} will lead to the series impedance. In~\pref{eq:mom}, the matrix $\matr{G}$ is the discretization of the  Green's function~\pref{eq:green}, and is made by $P \times P$ blocks $\matr{G}^{(p,q)}$ 
\begin{equation}
\matr{G} = 
\begin{bmatrix}
\matr{G}^{(1,1)} & \hdots & \matr{G}^{(1,P)}\\
\vdots & \ddots & \vdots \\
\matr{G}^{(P,1)} & \hdots & \matr{G}^{(P,P)}\\ 
\end{bmatrix}\,.
\end{equation}
The block $\matr{G}^{(p,q)}$ describes the contribution of the equivalent current on conductor $q$ to the vector potential on conductor $p$. The entries of $\matr{G}$ are given by a double integral involving the Green's function~\pref{eq:green}. This integral can be solved analytically with the approach we proposed in~\cite{tpwrd}. Analytic integration significantly reduced the CPU time needed to set up the matrix $\matr{G}$, which is dense, and makes the proposed algorithm very efficient. Indeed, all coefficient matrices in~\pref{eq:mom} can be computed analytically.
Finally, the matrix $\matr{U}$ in~\pref{eq:mom} follows from the relation between the conductor currents $\vect{I}$ and the equivalent current coefficients $\matr{J}$
\begin{equation}
\matr{I} = \matr{U}^{T} \matr{J}\,.
\label{eq:Udef}
\end{equation}
The matrix $\matr{U}$ has $P$ columns. In the $p$-th column, we have a ``1'' in the row corresponding to the position of $J_0^{(p)}$ in $\matr{J}$. If conductor $p$ is hollow, there is a ``1'' also in the row corresponding to $\widetilde{J}_0^{(p)}$. All other entries of $\matr{U}$ are zeros.

\subsection{Computation of the p.u.l. Impedance}

By combining~\pref{eq:mom} with~\pref{eq:sao} we finally obtain, with a few algebraic manipulations~\cite{tpwrd},  the p.u.l. resistance  and inductance
\begin{align}
\ppulpm{R}(\omega) & = \Re{\left [\transp{\matr{U}} (\matr{1} - j\omega \mu_o \matr{Y}_s \matr{G})^{-1} \matr{Y}_s \matr{U} \right ]^{-1}}\,,
\label{eq:ppulpR} \\
\ppulpm{L}(\omega) & = \omega^{-1} \Im{\left [\transp{\matr{U}} (\matr{1} - j\omega \mu_o \matr{Y}_s \matr{G})^{-1} \matr{Y}_s \matr{U} \right ]^{-1}}\,.
\label{eq:ppulpL}
\end{align}

\section{Ground Return}
\label{sec:ground}

We now show how we include the effect of lossy ground in the proposed technique. We decompose the series impedance $\ppulpm{Z}$ of a buried cable as
\begin{align}
\ppulpm{Z}  &= \left (\ppulpm{Z}_{c} + \ppulpm{Z}_g \right )+\Delta \ppulpm{Z}_{\rm prox} \,,
\label{eq:Zbreakdown}
\end{align}
where $\ppulpm{Z}_{c}$ and $\ppulpm{Z}_{g}$ denote, respectively, the contributions of the cables and of the ground evaluated neglecting proximity effects, which are instead represented by the term $\Delta \ppulpm{Z}_{prox}$. 

Conventional EMTP tools compute the series impedance of cables using analytical formulae which account for skin effect in both conductors and earth, but ignore any proximity effect \cite{Ame80}. Therefore, they only return the first two terms of~\pref{eq:Zbreakdown}. 

On the other hand, the proposed method estimates $\Delta \ppulpm{Z}_{prox}$ very accurately, but does not incorporate the effect of the ground return ($\ppulpm{Z}_g$) since it has been developed assuming a lossless medium around the conductors. In what follows, we show an easy approach~\cite{Gus95} which permits to properly include ground return in the proposed technique. 

We capitalize on the fact that with MoM-SO one can easily exclude proximity effects by setting $N_p = 0$ for all conductors. If we calculate the impedance matrix twice, with $N_p >0$ and with $N_p = 0$, we can estimate the contribution of proximity as
\begin{equation}
\Delta \ppulpm{Z}_{\rm prox} = \ppulpm{Z}_{\rm MoM-SO}(N_p>0) - \ppulpm{Z}_{\rm MoM-SO}(N_p=0) \,.
\label{eq:Zprox}
\end{equation}


Next, we calculate $\ppulpm{Z}=\ppulpm{Z}_{c}+\ppulpm{Z}_g$  using a conventional approach (ex: Cable Constants~\cite{Ame80,Wed73} or analytic formulas \cite{Vel10}) and add $\Delta \ppulpm{Z}_{prox}$ to the result according to~\pref{eq:Zbreakdown}. 
The proposed approach assumes that conductors and ground are separated by an infinitesimally-thin insulation layer. Since the thickness of the insulation layer in a real power cable is much smaller than skin depth in ground, it can be safely neglected in the computation of the cable impedance.

%
%


This simplified approach is valid as long as the penetration depth in ground is much larger than the distance between the conductors. To see this, consider the correction term $\Delta \ppulpm{Z}_{prox}$ in~\pref{eq:Zprox}. This correction is independent of the chosen return path (reference conductor)  provided that the same return is used in the two calculations ($N_p> 0$) and ($N_p=0$), and that the return path is far away from the conductors. This implies that one would get the same result if one had chosen the return path to be that of the classical ground return formula for $\ppulpm{Z}_g$ in~\pref{eq:Zbreakdown}.

%
%
%
\section{Numerical Results}
\label{sec:Results}
\subsection{Validation against Finite Elements}
\label{sec:result3shell}

\begin{figure}[t]
\centering
\includegraphics[width=.75\columnwidth, viewport= 150 285 450 490]{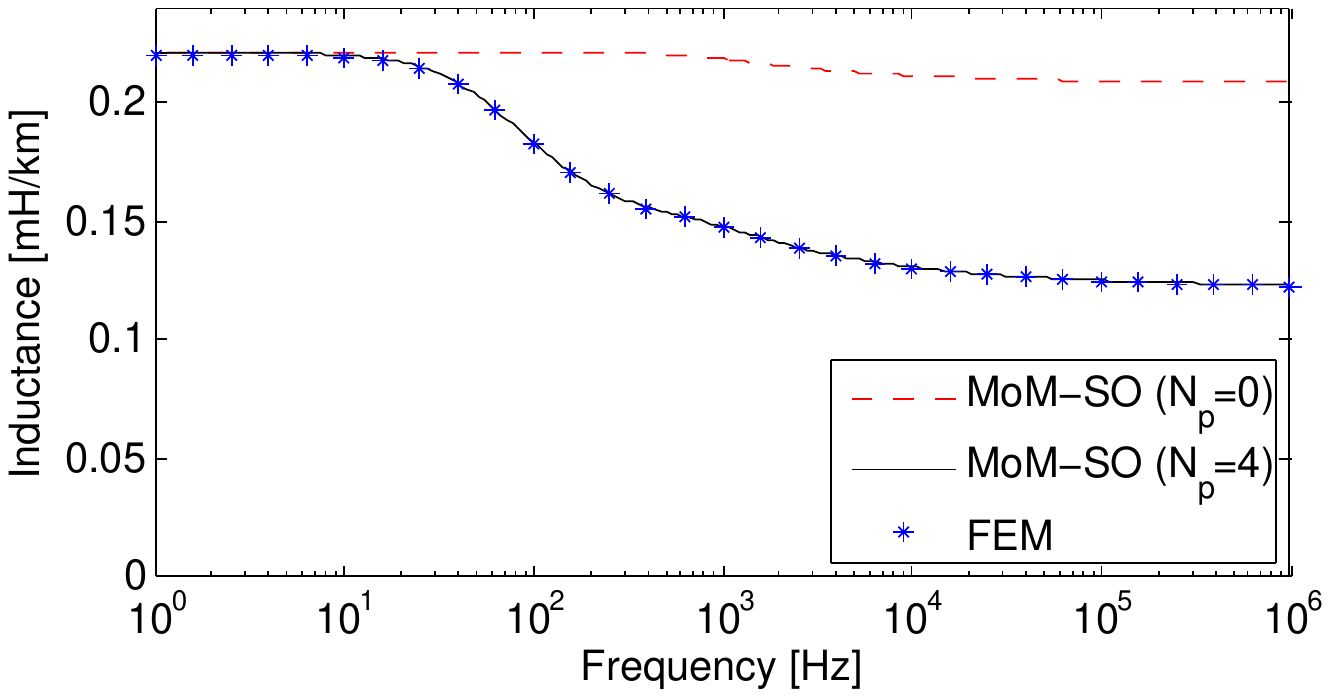}
\centering
\includegraphics[width=.75\columnwidth, viewport= 150 285 450 490]{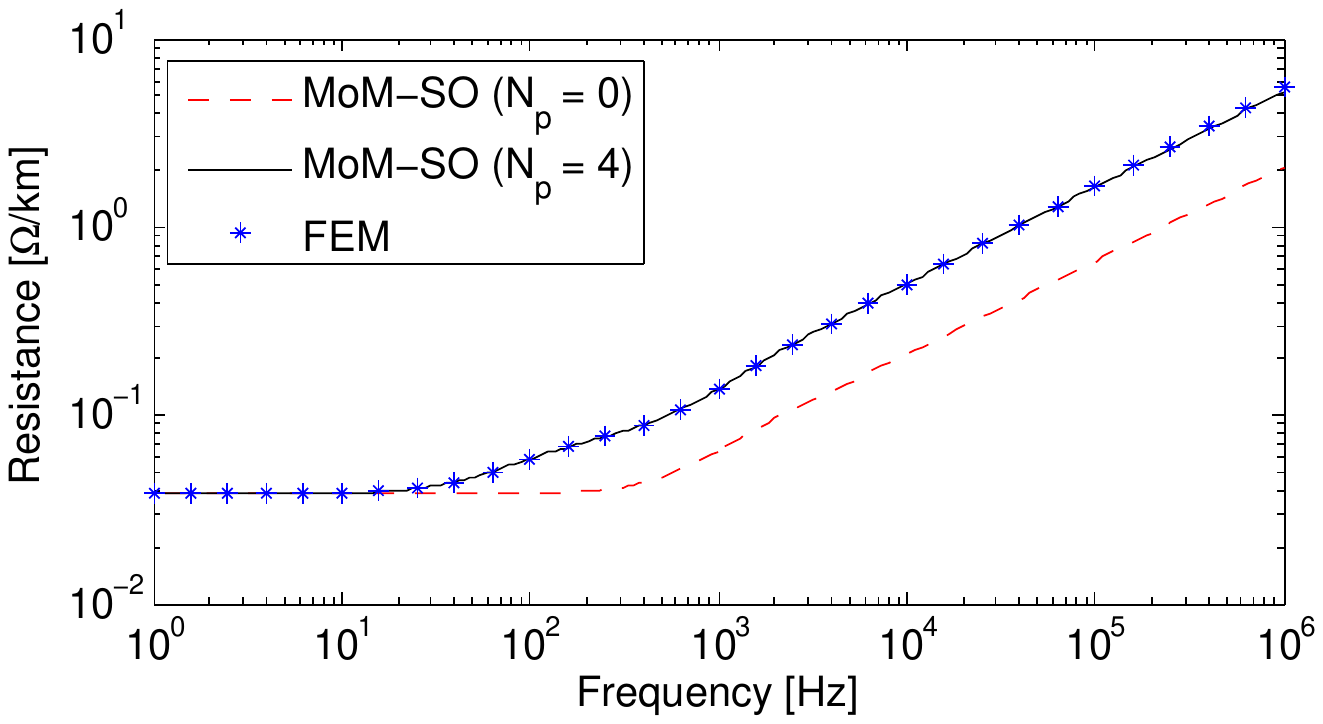}
\caption{Positive sequence inductance and resistance for the three coaxial shells system of Sec.~\ref{sec:result3shell}.}
\label{fig:InductanceResistance2}
\end{figure}

We first validate the proposed MoM-SO approach against FEM computation~\cite{B09} for a system of three uniformly-spaced coaxial shells surrounded by lossless medium. The center-to-center distance between the shells is $45~{\rm mm}$. Shells have a diameter of $40~{\rm mm}$ and thickness of $4~{\rm mm}$.
The conductivity of each shell is $58\cdot 10^6~{\rm S/m}$. We calculate the positive-sequence resistance and inductance using MoM-SO with orders $N_p=0$ (no proximity effects) and $N_p=4$ (with proximity effects). Result, illustrated in Fig.~\ref{fig:InductanceResistance2}, demonstrate an excellent agreement between MoM-SO and FEM. By comparing the two curves obtained with MoM-SO, one can appreciate the influence of proximity effects on the parameters of this cable, which becomes significant at medium/high frequency. Neglecting proximity leads to an overestimation of the series inductance, and of an underestimation of losses. 

Timing results, presented in Table~\ref{tab:Timing}, show that the proposed method is 34 times faster then FEM. This remarkable speed up arises from two differences between MoM-SO and FEM:
\begin{itemize}
\item finite element methods have to mesh the entire cross-section of the cable, instead of the sole surface which is sufficient for MoM-SO. This difference is particularly significant at high frequency, where the small skin depth imposes a very fine mesh in FEM;
\item MoM-SO uses very few unknowns per conductor. For example, when $N_p = 4$, the field/current Fourier series have only 9 coefficients for solid conductors, and 18 for hollow conductors.
\end{itemize}
%

\begin{table}[t]
\centering
\caption{Timing results for the three coaxial shells example discussed in Sec.~\ref{sec:result3shell}.}
\begin{tabular}{| p{3.9cm} | c | c | c |}
\hline
& \multicolumn{2}{|c|}{\bf Proposed (MoM-SO)} & \\ 
\cline{2-3}
& ${N_p=0}$ & $ N_p=4$ & \bf{FEM} \\ \hline
Computation of $\matr{G}$ & 0.011~s & 0.254~s &  \\ \hline
Per-frequency impedance calculation  & 0.025~s & 0.040~s & 1.583~s* \\ \hline
\end{tabular} \\
\label{tab:Timing} 
{All computations were performed on a system  with a 2.5 GHz CPU and 16~GB of memory.\\
*Positive sequence only. Mesh size: 41,562 triangles.}
\end{table}


\subsection{Validation of Ground Return}
\label{sec:resultground}
%

We demonstrate the adequacy of the approach proposed in Sec.~\ref{sec:ground} for the inclusion of ground return by a direct comparison against a FEM computation~\cite{B09}. We consider two close conductors that are buried in an infinite earth with $\sigma_o =0.1{~\rm S/m}$. The radius of each conductor is $a = 25{\rm~mm}$, while separation is $D = 70{\rm~mm} $. The conductivity of each conductor is $\sigma = 58\cdot 10^5 {~\rm S/m}$. We wish to calculate the impedance matrix $\ppulpm{Z}$ at $10{\rm~kHz}$. The impedance by the classical approach \cite{Vel10} is obtained as
\begin{equation}
\ppulpm{Z}_{\rm analytic} = 
\begin{bmatrix}
{\ppulp Z}_1 + {\ppulp Z}_{g,s} & {\ppulp Z}_{g,m} \\
{\ppulp Z}_{g,m} & {\ppulp Z}_1 + {\ppulp Z}_{g,s} 
\end{bmatrix} \,,
\label{eq:2condZ}
\end{equation}
where
\begin{align}
{\ppulp Z}_1 &= \frac{m}{2\pi a \sigma} \frac{{\cal I}_0(ma)}{{\cal I}_1(ma)}\,, \\
{\ppulp Z}_{g,s} &= \frac{m_o}{2 \pi a \sigma_o} \frac{{\cal L}_0(m_o a)}{{\cal L}_1(m_o a)}\,, \\
{\ppulp Z}_{g,m} &= \frac{m_o}{2\pi a^2 \sigma_o} \frac{{\cal L}_0(m_o D)}{({\cal L}_1(m_o a) )^2} \,,
\end{align} 
for $ m = \sqrt{j\omega \mu \sigma} \,, $  $m_o = \sqrt{j\omega \mu_o \sigma_o} \,,$
and where ${\cal I}_n(.)$ and ${\cal L}_n(.)$ are the modified Bessel functions of first and second kind~\cite{Abr64} of order $n$.

When calculating the correction~\pref{eq:Zprox} using MoM-SO, we use as return a tubular conductor of 10-m radius and 1-mm wall thickness with $\sigma =58 \cdot 10^5 {\,\rm S/m}$. 

The result is validated against a FEM computation~\cite{B09}. Since the penetration depth in earth is $\delta=15.9 {\rm~m}$ at the given frequency and soil resistivity, it is sufficient to use a boundary of radius $3\delta=48 {\rm~m}$. 

Table~\ref{tab:Zground} shows the impedance values calculated in the various steps, presented in the form of common mode and per-phase loop impedances.
It is observed that the simplified approach agrees with the FEM result with an error smaller than 0.1\% for both the real and imaginary part of \pref{eq:2condZ}. 

\begin{table}
\caption{Example of Sec~\ref{sec:resultground}: common mode and loop mode p.u.l. impedance at $10\, {\rm kHz}$ in ${\rm \Omega /km}$. The difference between the proposed approach~\pref{eq:Zbreakdown} and the analytic formula~\pref{eq:2condZ} is the contribution of proximity.}
\begin{tabular}{|p{1.8cm}|c|c|c|c|}
\hline
& \multicolumn{2}{|c|}{Common Mode} & \multicolumn{2}{|c|} {Loop Mode}\\ \hline
& Real  & Imaginary  & Real  & Imaginary  \\ \hline
Proposed~\pref{eq:Zbreakdown} & 20.39 & 142.68 & 0.75 & 11.64 \\ \hline
FEM & 20.38 & 142.67 & 0.75 & 11.64 \\ \hline
Error between~\pref{eq:Zbreakdown} and FEM & 0.0589\% & 0.00640 \% & 0.0818 \% & 0.0350 \% \\ \hline
Error between~\pref{eq:Zbreakdown} and~\pref{eq:2condZ} & -0.48\% & 0.97\% & -26.92\% & 15.67\% \\ \hline
\end{tabular}
\label{tab:Zground}
\end{table}

\begin{figure}[t]
\centering
\includegraphics[width=.75\columnwidth, viewport= 150 285 450 490]{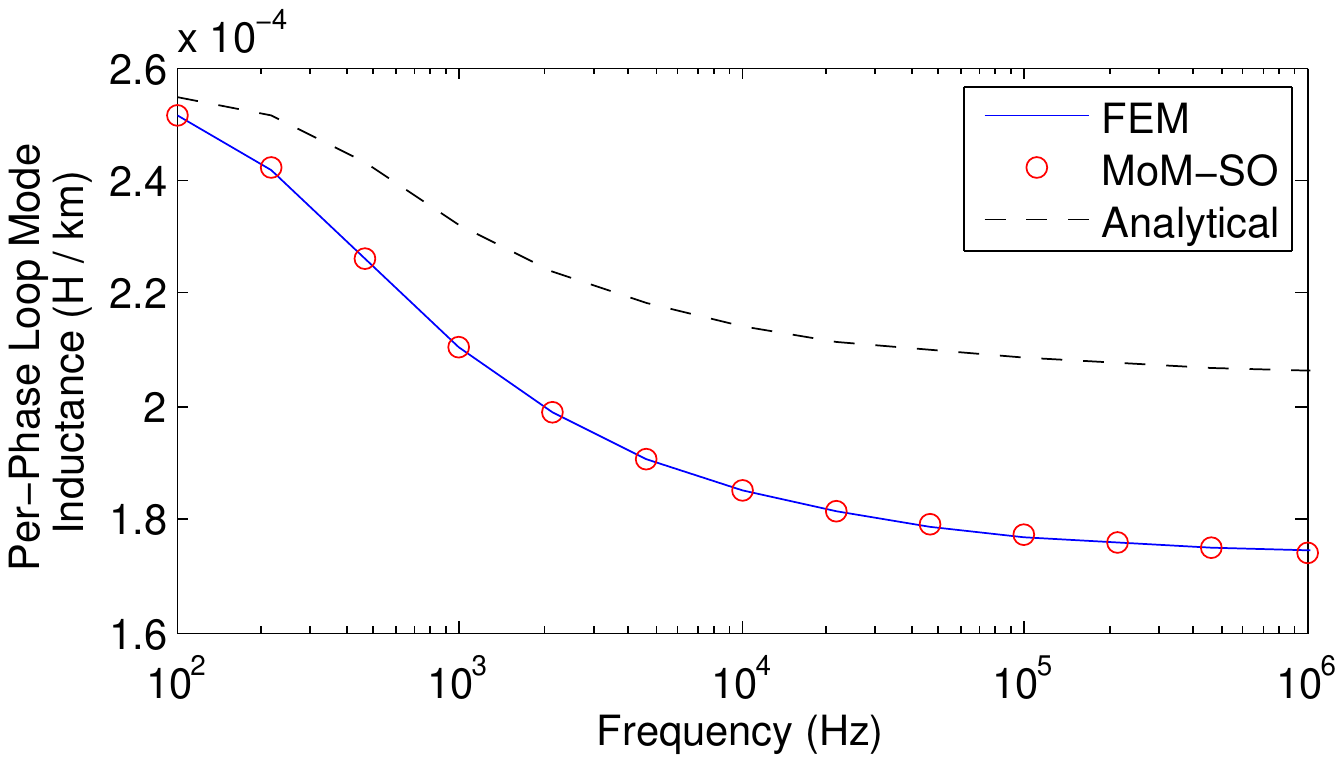}

\centering
\includegraphics[width=.75\columnwidth, viewport= 150 285 450 490]{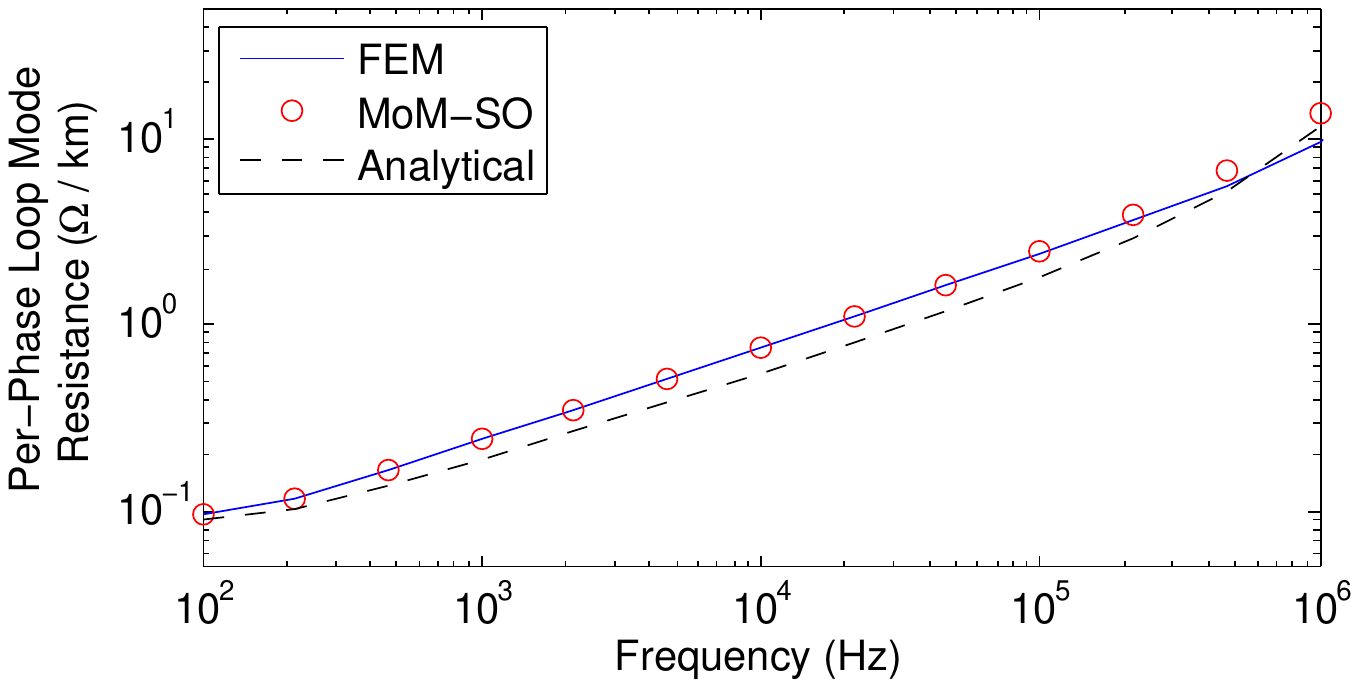}

\caption{System of Sec.~\ref{sec:resultground}: loop mode inductance and resistance over frequency computed with the proposed approach~\pref{eq:Zbreakdown}, the analytic formula~\pref{eq:2condZ} and FEM. The difference between the curve obtained with the proposed approach and with analytic formulas is the contribution of proximity.}
\label{fig:Inductance3}
\end{figure}

The per-phase loop-mode inductance and resistance are plotted over frequency in Fig.~\ref{fig:Inductance3}. The results validate the proposed approach for ground return inclusion, since the obtained results match closely those computed with FEM. A similar agreement was obtained for the common mode inductance and resistance.

\subsection{Transient Overvoltages in a Crossbonded Cable System}
\label{sec:Ex3}
\subsubsection{Cable Data}

\begin{figure}[!t]
\centering
\includegraphics[width=0.8\columnwidth, viewport= 0 20 350 100]{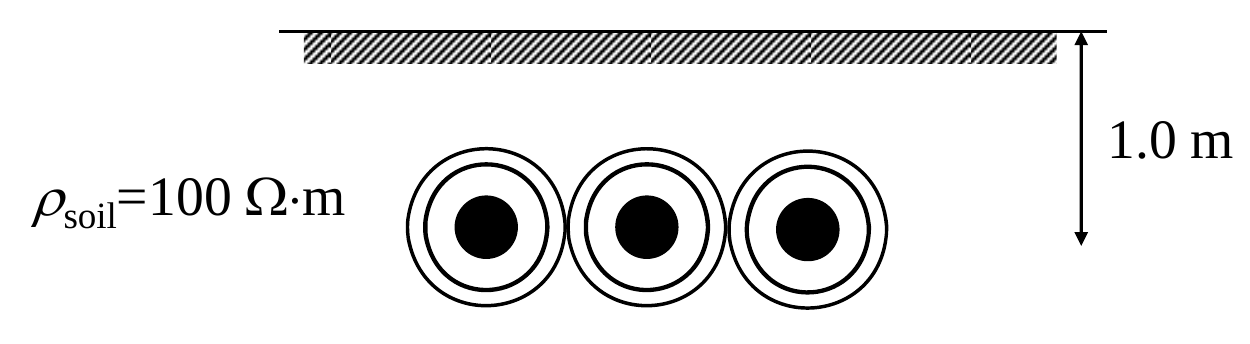}
\caption{Underground single core cables considered in Sec.~\ref{sec:Ex3}.}
\label{fig:ThreeSCCables}
\end{figure}

We consider the modeling of three single core cables buried in a homogeneous soil as shown in Fig.~\ref{fig:ThreeSCCables}. The cables are touching, leading to a significant proximity effect for waves that propagate external to the sheaths. The cable geometry and material properties are listed in Table~\ref{tab:SCparameter}.

We compute the series impedance matrix $\ppulpm{Z}$ in two alternative ways: using Wedepohl's analytical approach which considers skin effects and ground return~\cite{Wed73}, and using MoM-SO. With MoM-SO, we use as reference conductor a tubular conductor of 10-m radius, 1-mm thickness and conductivity equal to that of the core conductor. The shunt admittance $\ppulpm{Y}$ is established by standard analytical formulas~\cite{Wed73}. In both cases, we evaluate the impedance at 120 logarithmically spaced points distributed from $1~{\rm Hz}$ to $1~{\rm MHz}$.

\begin{table}
\centering
\caption{Single core cables of Sec.~\ref{sec:Ex3}: geometrical and material parameters.}
\begin{tabular}{|c|c|}
\hline
Core & Outer diameter = 39~mm, $\rho = 3.365\cdot 10^{-8} {\rm~\Omega \cdot m}$ \\ \hline
Insulation & $t = 18.25 {\rm~mm} $, $\epsilon_r = 2.85$ \\ \hline
Sheath & $t = 0.22{\rm~mm}$, $\rho = 1.718\cdot 10^{-8} {\rm~\Omega \cdot m}$ \\ \hline
Jacket & $t = 4.53{\rm~mm}$, $\epsilon_r = 2.51$ \label{tab:SCparameter} \\ \hline
\end{tabular}
\end{table}

\subsubsection{Timing Results}
 
Timing results are presented in Table~\ref{tab:TimingEx3} for the two runs of MoM-SO necessary to evaluate~\ref{eq:Zprox} and consequently~\pref{eq:Zbreakdown}. In total, for computing the cable impedance at 120 frequency points, the proposed approach takes 
less than 10~s. This result confirms the efficiency of MoM-SO, that can provide, in a few seconds, cable parameters with the accuracy of a FEM simulation.
 

\begin{table}[t]
\centering
\caption{Timing results for the example of Sec.~\ref{sec:Ex3}.}
\begin{tabular}{| p{3.0cm} | c | c |}
\hline
& MoM-SO: ${N_p=0}$ & MoM-SO: $ N_p=4$ \\ \hline
Computation of $\matr{G}$ & 0.019~s & 0.608~s  \\ \hline
Impedance computation (per frequency sample) & 0.030~s & 0.047~s \\ \hline
\end{tabular}
\label{tab:TimingEx3} 
\end{table}

\subsubsection{Modal Analysis}
\label{sec:Ex3ModalAnalysis}
\begin{figure}[t]
\centering
\includegraphics[width=.75\columnwidth, viewport= 150 290 450 490]{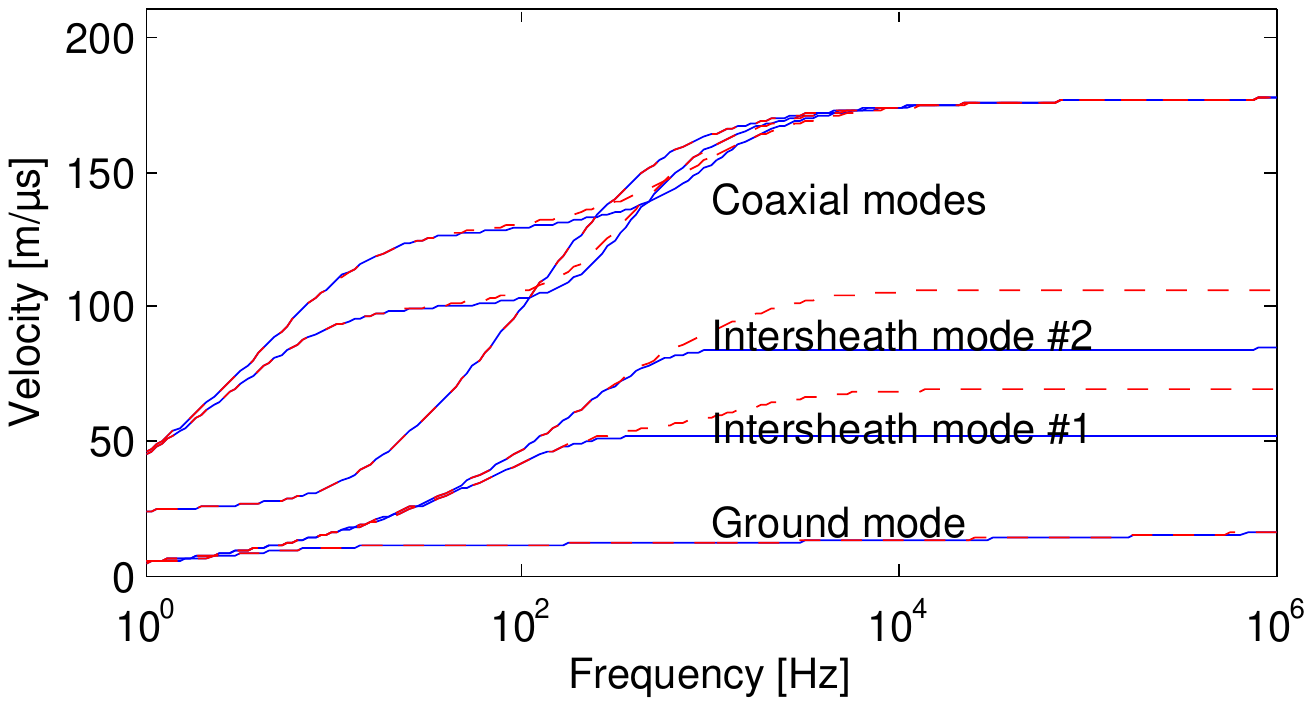}
\caption{Single core cables of Sec.~\ref{sec:Ex3ModalAnalysis}: velocity of different modes calculated from the series impedance obtained with analytic formulas (solid blue lines) and the proposed approach which includes skin, proximity, and ground return effects (red dashed lines).} 
\label{fig:ModalVelocity2}
\end{figure}

Figure~\ref{fig:ModalVelocity2} compares the modal velocities of propagation obtained when $\ppulpm{Z}$ is computed with and without the inclusion of proximity effects. Analytical approach considers skin effect but neglects proximity effect. To capture the proximity effect we use MoM-SO with order $N_p=4$. By combining the MoM-SO results with Wedepohl's analytical formulas~\cite{Wed73}, we get by~\pref{eq:Zbreakdown} an impedance matrix which accounts for skin, proximity and earth return effects. Clearly, proximity effect increases the propagation speed of the intersheath waves.

%
%
%

\subsubsection{Modeling for Transient Calculations}
\label{sec:Ex3TransientMat}
We use the computed series impedance to perform a transient simulation. For this purpose, we calculate the parameters of the Universal Line Model~\cite{Mor99,Gus99} using the series impedance $\ppulpm{Z}$ and shunt admittance $\ppulpm{Y}$ found in the previous section. The model is formulated in terms of the phase-domain characteristic admittance $\ppulpm{Y}_c$ and propagation matrix $\matr{H}$. We used 12 poles for the fitting of $\ppulpm{Y}_c$ and 14 poles for fitting each of the four modal delay groups of $\matr{H}$.   

%


\subsubsection{Transient Overvoltages}
\label{sec:Ex3TransientOvervoltages}
\begin{figure}[!t]
\centering
\includegraphics[width=0.7\columnwidth, viewport= 0 0 280 130]{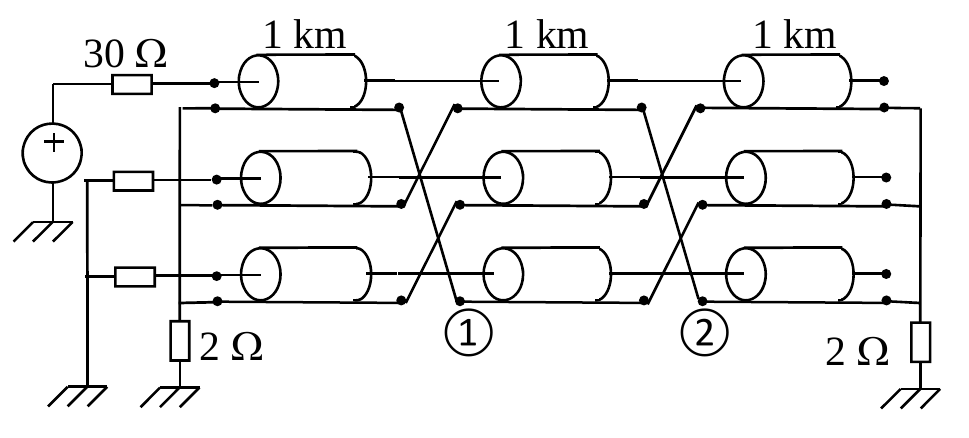}
\caption{Configuration of the crossbonded cable considered in Sec.~\ref{sec:Ex3TransientOvervoltages}. A unit step voltage is applied at one end of the cable.}
\label{fig:setup}
\end{figure}

\begin{figure}[!t]
\centering
\includegraphics[width=.75\columnwidth, viewport= 150 295 450 490]{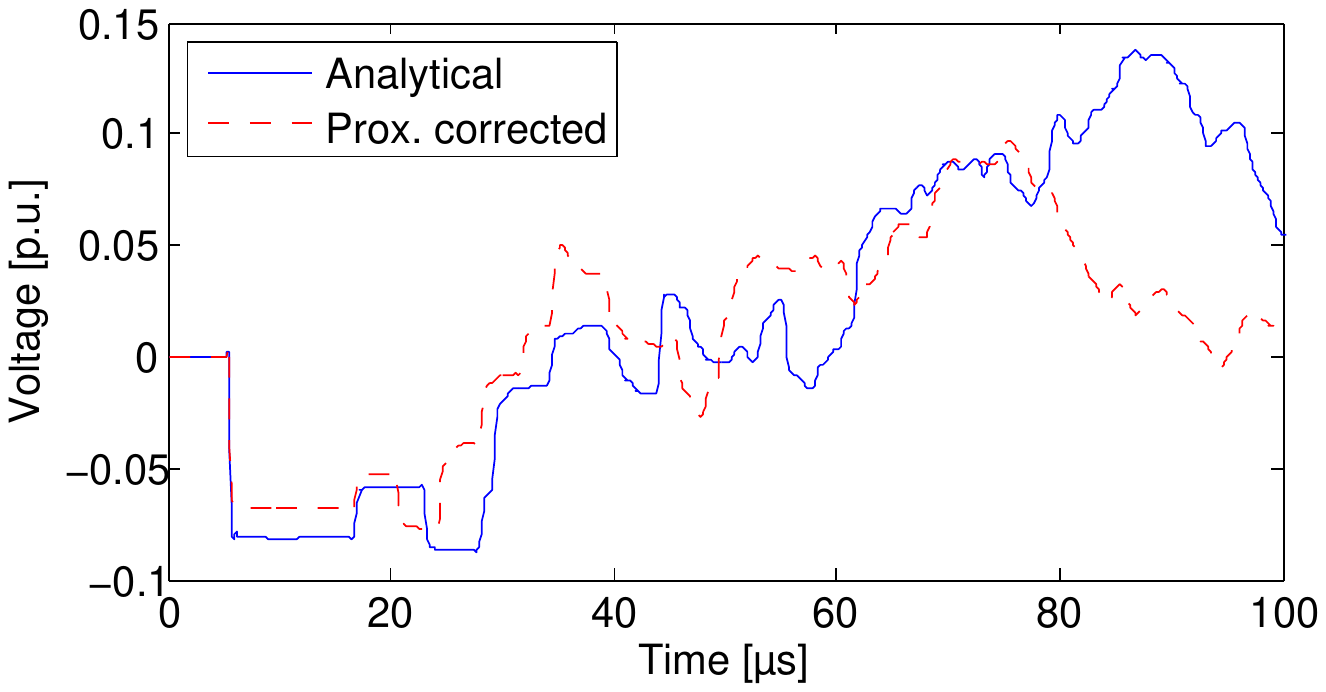}
\caption{Overvoltage at node \# 1 of the crossbonded cable of Fig.~\ref{fig:setup}. The voltage has been computed with the series impedance obtained with the proposed technique, which accounts for proximity effects, and with the analytical formulas of~\cite{Wed73}, which neglect proximity.}
\label{fig:Transient1}
\end{figure}

We wish to simulate transient overvoltages within a major section of a cross-bonded cable system. The obtained cable model was exported to the PSCAD simulation tool~\cite{PSCAD} and utilized in a transient simulation with crossbondings and terminal conditions as shown in Fig.~\ref{fig:setup}. The simulation was done with a unit step voltage excitation. Figs.~\ref{fig:Transient1} show the simulation results for the sheath voltage at nodes \#1 in Fig.~\ref{fig:setup}. The result is shown when $\ppulpm{Z}$ has been obtained with the analytical approach~\cite{Wed73}, which neglects proximity effects, and with the proposed approach. Clearly, the proximity effect has a very strong impact on the voltage waveforms.

%


\section{Discussion}

In Section~\ref{sec:Results}, we demonstrated the extended MoM-SO method for the modeling of a typical cable system that includes three solid phase conductors and three tubular screens, as well as earth return. In order to compute the 120 samples required for this system, the proposed approach took about 10~s. Although this is slower than standard analytical approaches, it is in our opinion fast enough to be effectively used in EMTP-type tools, in particular when proximity effects are suspected to be of concern.

We have also applied the new MoM-SO method for the modeling of pipe-type cables with similar results \cite{ipst2013}. Again, proximity effect was found to have a strong influence on the computed series impedance, which was accurately captured by the proposed approach with an acceptable CPU time.

\section{Conclusion}

The MoM-SO approach is an efficient method to compute the series impedance of power cables including skin and proximity effects. In this paper, we extended the methodology to hollow round conductors, useful to efficiently represent coaxial screens and armouring structures present in pipe-type cables. We have also shown how the influence of lossy ground can be taken into account. Compared to other proximity-aware techniques, such as finite elements, the proposed method is much faster, thanks to a surface-based formulation. Finally, the method has been used to predict a transient overvoltage in a cross-bonded cable system. The obtained results validate the technique and remark on the importance of accounting for proximity effects in cables with closely-space conductors.

\clearpage
\bibliography{biblio}
\bibliographystyle{plain}

\end{document}